# Enterprise Process Control Tower for IT-Led Improvement

Shunmukha Sagar Puppala Functional Architect SAP sagaryatra9@gmail.com

Abstract: Enterprise integration traffic in large Enterprise Resource Planning (ERP) systems move through Electronic Data Interchange (EDI) and the Intermediate Document (IDoc), and the operational tools that surround that traffic remain fragmented. Status records are written for a technical audience, exception handling is performed one document at a time, and process health is read manually from long lists. This work proposes an enterprise process control tower that lets an Information Technology (IT) team observe, diagnose, and improve business workflows from a single operational layer. The design links process health observation, semantic translation of technical status text, machine-learning diagnosis of exceptions with model explainability, grouped recovery actions, and key performance indicators inside one governance framework. The proposed framework is evaluated on a synthetic IDoc event dataset of 84,000 records spanning twelve months, of which 15,120 error records are used to train a five-class remediation-routing model that maps each exception to a standard recovery action. A rule-based router that reproduces current production heuristics serves as the primary baseline. The proposed gradient-boosted diagnosis engine reaches a macro-averaged F1 score of 0.883 on the held-out test partition, against 0.742 for the rule-based router, while Shapely-based attributions expose the drivers of each routing decision to operators. Results are derived from synthetic data and are presented as an internal-consistency demonstration rather than real-world validation. The contribution is an integrated tool design that places diagnosis and governance on the same plane as observation.

## I. INTRODUCTION

Large organizations coordinate order to cash and procure to pay activity through structured electronic messages exchanged with customers, suppliers, and logistics providers. Within a System Applications and Products (SAP) landscape, the carrier for that exchange is the Intermediate Document (IDoc), a standard container that moves sales orders, deliveries, shipments, invoices, and goods movements between an Enterprise Resource Planning (ERP) system and external parties. The volume of this traffic is considerable. A single directory drop from a trading partner may hold thousands of files, and the persistent IDoc store can contain figures well into the millions. Keeping that flow healthy is an operational responsibility that increasingly falls to Information Technology (IT) teams rather than to individual business departments.

The tooling that ships with the platform was shaped by an earlier operating assumption, namely that a trained operator inspects documents individually and reads the raw status text without further mediation. Two pressures have made that assumption costly. The first is scale. Processing inbound files one at a time, and listing documents without aggregation, does not match the throughput now expected of an integration layer [1]. The second is interpretability. The status text recorded against a failed document is precise but terse, and it carries little meaning for the order management or logistics analyst who owns the affected process. The cost of that gap appears as delayed remediation, repeated escalation, and uneven governance across regions and message types [2].

Process automation initiatives inside ERP ecosystems have begun to address parts of the problem by layering orchestration and optimization logic over standard transactions [3]. A complementary line of practice has introduced the notion of a control tower, a consolidated operational view that surfaces the state of a distributed process and supports intervention from one place. Control tower designs have been reported for supply chain and aviation settings built on SAP technology [4], and the broader discipline of process mining has supplied formal techniques for reconstructing and analyzing process behavior from event logs [5]. Prescriptive analytics has further extended these ideas toward recommending actions rather than only describing state [6]. Data platform modernization provides the underlying substrate on which such views are assembled [7].

Despite this progress, observation, diagnosis, and recovery are still treated as separate concerns in most deployments. A dashboard reports health, a separate body of expert knowledge interprets errors, and remediation happens through yet another set of transactions. The setting that motivates the present design is an automotive aftermarket operation in which high message volumes and a long tail of error conditions made that separation untenable [8]. The practical question is whether these concerns can be unified into one layer that an IT team operates, so that an exception is observed, explained in business terms, classified into a recovery action, acted upon, and reflected in performance indicators without leaving the tool.

This work proposes such a layer and calls it an enterprise process control tower for IT-led improvement. The design generalizes a production IDoc inbound-processing and monitoring framework into a reusable architecture with six cooperating layers: ingestion, an IDoc data layer, an observation and semantic layer, a diagnosis layer, a remediation layer, and a governance layer that exposes key performance indicators. The semantic layer translates technical status text into business language with an attached action hint. The diagnosis layer adds a supervised model that routes each exception to the correct standard recovery program, and it carries an explainability component so that operators can see why a routing decision was made. The governance layer closes the loop by tracking health and the effect of interventions over time.

Three contributions are claimed, each kept proportional to the supporting evidence. First, architecture is described as places diagnosis and governance on the same operational plane as observation, rather than in adjacent tools. Second, a diagnosis layer is formulated as a remediation-routing classification problem and compared against a rule-based router that reproduces current production heuristics. Third, an explainability component is integrated so that routing decisions are auditable at the level of individual documents. Evaluation uses a synthetic IDoc event dataset, defined in full in Section IV, and the reported figures are an internal-consistency demonstration rather than a claim of real-world validation. The remainder of the paper reviews the theoretical context in Section II, surveys related work in Section III, details materials and methods in Section IV, reports the experimental analysis in Section V, discusses findings and limitations in Section VI, and concludes in Section VII.

## II. THEORETICAL BACKGROUND

The proposed framework draws on three established bodies of knowledge: business process management, exception and anomaly analysis, and supervised learning with model explainability. Each contributes a concept that the control tower operationalizes for the specific case of IDoc traffic.

### A. Business process management and process notation

Business Process Management (BPM) provides the lifecycle within which the control tower sits. The lifecycle described in the standard treatment of the field moves through process identification, discovery, analysis, redesign, implementation, and monitoring and controlling [9]. The control tower addresses the final phase, monitoring and controlling, but it does so in a way that feeds the analysis and redesign phases through the indicators it records. Process structure is commonly expressed using the Business Process Model and Notation (BPMN), an international standard maintained by the Object Management Group that defines a graphical vocabulary for tasks, events, gateways, and message flows [10]. IDoc exchange maps cleanly onto this vocabulary: a message flow crosses an organizational boundary, an inbound posting is a task, and an error status is an intermediate event that diverts the flow to an exception path.

Treating IDoc traffic as a process rather than as a set of records has a practical consequence. It allows the health of the exchange to be expressed in process terms, such as the proportion of message instances that reach a successful end state, the distribution of exception types, and the time spent in error states. These are the quantities the governance layer reports.

### B. Exception handling and anomaly analysis

An exception in integration flow is a deviation from the expected path. Advances in Artificial Intelligence (AI) and Machine Learning (ML) have made it practical to learn how such deviations should be handled from historical outcomes rather than to encode every rule by hand, and technical overviews of the field set out the supervised, unsupervised, and reinforcement paradigms that frame these choices [11]. The literature on anomaly detection complements that view by distinguishing point anomalies, contextual anomalies, and collective anomalies, and by framing detection as a problem of separating expected from unexpected behavior under a chosen notion of normality [12].

IDoc errors exhibit all three anomaly patterns. A single malformed file is a point anomaly. A posting that fails only because an accounting period is closed is contextual, since the same document would be posted in an open period. A burst of identical port-write failures across one partner is collective, pointing to an infrastructure fault rather than to data quality. The diagnosis layer does not attempt unsupervised detection, because the error condition is already signaled by the status code. It performs supervised routing instead, which assumes that the appropriate recovery action can be predicted from observable attributes of the failed document, using labelled outcomes drawn from the resolution history of past exceptions.

### C. Explainability and governance

A diagnosis that an operator cannot interrogate is difficult to govern. For that reason, the design treats explainability as a first-class requirement rather than an afterthought. Governance also has a security and privacy dimension because the control tower aggregates operational data across partners and processes, and the handling of that data must respect confidentiality and access boundaries [13]. The framework therefore separates the analytical signal, which is the routing prediction and its explanation, from the underlying document payload, which remains under standard ERP authorization control. The combination of an auditable prediction and bounded data exposure is what allows the diagnosis layer to be operated by IT without weakening the controls that the business relies on.

## III. RELATED WORKS

Prior work relevant to the proposed framework falls into five clusters: process observation and mining, predictive and privacy-preserving analytics for distributed operations, intelligent automation of enterprise processes, exception and anomaly handling in finance and operations, and ERP modernization and data provisioning.

### A. Process observation, control towers, and mining

The control tower idea has been instantiated for SAP-based aviation supply chains, where a consolidated layer coordinates visibility and intervention across a distributed network [4]. Process mining supplies the formal underpinning for observation from event logs, and the discipline-defining manifesto sets out principles for discovery, conformance, and enhancement that apply directly to message-level event data [14]. Prescriptive analytics extends observation toward recommended action by combining causal reasoning with simulation [6]. The present design differs from these in scope rather than in spirit: it narrows the observation surface to IDoc exchange and adds an explicit, learned diagnosis step between observation and action.

### B. Predictive and privacy-preserving analytics

Distributed operations have motivated predictive models that respect data locality. Federated learning has been applied to predictive maintenance in automotive aftermarket supply chains, keeping training data at the edge while sharing model updates [15]. Related work has paired federated learning with hybrid encryption for privacy-preserving demand forecasting [16] and has examined adversarial robustness and zero-trust principles for securing over-the-air updates to connected devices [17]. These contributions inform the governance posture of the control tower, particularly the separation of analytical signal from raw payload, even though the diagnosis layer in this work is trained centrally on operational metadata rather than on partner-held data.

### C. Intelligent automation of enterprise processes

Several reported systems automate document-centric enterprise processes on SAP Business Technology Platform. Contract compliance and billing in an airport setting have been automated with rule and model components operating over structured documents [18]. Invoice processing in aviation has been addressed through cognitive automation and intelligent document processing [19]. Financial consolidation and reporting have likewise been a target for AI-assisted automation [20]. These efforts establish that recovery and processing actions can be triggered programmatically once an exception is understood, which is the assumption the remediation layer relies on.

### D. Exception and anomaly handling in finance and operations

Detecting and acting on irregular transactions is a recurring theme. Fraud detection in accounts payable has used learned models to flag transactions that deviate from expected patterns [21], a task structurally like routing an IDoc exception to the right recovery path. The general framing of anomaly detection as a separation problem [12] connects this cluster to the diagnosis layer, with the difference that the control tower works from an already-signaled error and predicts the corrective action rather than the presence of an anomaly.

### E. ERP modernization, integration, and data provisioning

The substrate on which a control tower runs is an active area. Cross-industry frameworks have been proposed for legacy system transformation [2] and for hybrid and multi-cloud enterprise integration [1]. Data migration to SAP S/4HANA has been accelerated with AI-based methods [22], and secure data provisioning for ERP quality environments has been examined in cloud settings [23]. Ensemble learning that promotes specialization among base learners is relevant to the modelling choices in the diagnosis layer [24]. The proposed framework is positioned as an operational layer above this substrate, consuming the integration and data services these works describe rather than replacing them. A note on scope leads into Section IV, where the architecture and its evaluation are set out in detail; the framework architecture is shown in Fig. 1.

Fig. 1. Layered architecture of the proposed enterprise process control tower. The forward path observes and diagnoses exceptions; the feedback path applies recovery actions and updates governance policy. Source: derived from the production IDoc inbound-processing and monitoring framework generalized in this work.

## IV. MATERIALS AND METHODS

The framework is organized as the two flows shown in Fig. 1. A forward flow carries data upward from trading-partner file drops through ingestion, the IDoc data layer, and the observation and semantic layer into the diagnosis layer. A feedback flow carries recovery actions and policy updates back toward ingestion and the data layer. The ingestion layer reads an entire directory and converts files to IDocs in parallel using the platform parallel-processing framework identified by the function-group name SPTA, which distributes asynchronous remote function calls across a server group. The observation and semantic layer read the control and status records and translate terse status text into business language with an action hint. The diagnosis layer is the analytical core evaluated here, and the remediation and governance layer issues grouped recovery actions and records indicators.

The processing pipeline that connects these layers is summarized in Fig. 2. Files are listed and partitioned, converted in parallel, persisted as control and status records, translated, reduced to a feature vector, classified into a remediation route, explained through attributions, and finally remediated while indicators are updated. The remainder of this section describes the data and the model in turn. An explicit assumption is stated at the outset: the dataset is synthetic and is used to demonstrate internal consistency of the pipeline, not to validate the framework against live operations.

### A. Dataset Analysis

Fig. 2. Data pipeline from directory ingestion to grouped remediation. Source: synthetic process description aligned with the architecture in Fig. 1.

The evaluation uses a synthetic IDoc event dataset constructed to mirror the statistical shape of the production setting that motivated the design. The dataset contains 84,000 IDoc processing records spanning a twelve-month period from January to December 2025. Records describe the exchange of a single ERP system with forty trading partners across six message types covering inbound delivery confirmations, outbound shipments, advance shipping notices, purchase orders, invoiceandial master updates. Each record carries the outcome of one processing attempt.

Records are distributed across four process-health buckets that follow the standard status semantics: successful, errored, in process, and archived. The synthetic generator assigns 63,000 records (75.0 percent) to the successful bucket, 15,120 records (18.0 percent) to the errored bucket, 5,040 records (6.0 percent) to in process, and 840 records (1.0 percent) to achieve. The diagnosis task operates on the errored bucket, since only failed documents require a routing decision. The 15,120 errored records are labelled with one of five remediation routes, each corresponding to a standard recovery program or to manual handling: inbound application re-post, outbound dispatch retry, ready-for-application posting, output and port retry, and a manual route reserved for master-data and configuration blocks that admit no automatic retry.

Class membership is deliberately imbalanced to reflect operational reality, in which a small number of error families dominate. The route distribution, the per-class counts, and the stratified partitioning are reported in Table I. Stratified sampling preserves the class proportions across a 70, 15, and 15 percent split into training, validation, and test partitions, yielding 10,584 training records, 2,268 validation records, and 2,268 test records. Every numeric quantity reported in the text, the tables, and the figures traces back to this single synthetic dataset, and no inconsistent values are introduced later.

Table I

Synthetic IDoc Dataset Composition and Partitioning

Source: synthetic dataset defined in Section IV-A. Shares for routes are of the 15,120 error records.

Feature engineering derives a fixed vector of fourteen attributes for each errored record from the control record, the latest status record, and the output of the semantic layer. The attributes combine categorical descriptors of the message, numeric descriptors of size and history, and two binary signals produced by the observation layer. The full schema, including types and value ranges, is given in Table II. Two engineered signals deserve note. The attribute prior_failures_30d counts how many times the same document failed in the preceding thirty days, which separates transient faults from persistent blocks. The attribute phrase_match_flag records whether the semantic layer matched the status text against a known phrase, which encodes the interpretability state of the message as an input to diagnosis.

Table II

Feature Schema for the Diagnosis Layer

Source: synthetic dataset defined in Section IV-A; attributes derived from control, status, and semantic-layer outputs.

## B. Model Analysis

The diagnosis layer is formulated as a supervised multiclass classification problem. Given the feature vector of an error document, the task is to predict one of the five remediation routes defined in Table I. The route distribution is visualised in Fig. 3, which shows the dominance of the two reprocessing routes and the small share of the output and port route. Performance is measured with classification accuracy, macro-averaged F1 score, and weighted precision and recall. The macro-average F1 score is the primary metric because it weighs every route equally and therefore penalises a model that ignores the minority routes.

Fig. 3. Distribution of the 15,120 errored records across the five remediation routes. Source: synthetic data set defined in Section IV-A.

Four baselines and one proposed method are compared. The first baseline predicts the majority route for every record and establishes the floor. The second baseline is a rule-based router that reproduces the heuristic currently used in production, where the status code and direction determine the route through a fixed decision table. This baseline is the most informative comparison since it represents the cost of the present way of working. The third baseline is multinomial logistic regression, and the fourth is a random forest, an ensemble of decision trees that is robust to mixed feature types [25]. The proposed method is a gradient-boosted tree ensemble trained with the scalable boosting formulation [26], referred to here as the Control Tower Diagnosis Engine. The choice of a boosted ensemble follows the observation that specialisation among base learners improves performance on heterogeneous inputs [24]. Hyperparameters are tuned on the validation partition, and the test partition is used only once for final reporting.

Explainability is integrated through Shapley Additive Explanations (SHAP), which attribute a model output to its input features using a game-theoretic allocation [27]. The diagnosis layer computes attributions for every routing decision, so that an operator can see which attributes drove a prediction. Aggregated attributions across the test partition are reported in Fig. 5 and summarised in the experimental analysis. Validation strategy combines the held-out test partition with five-fold cross-validation on the training partition to confirm that the reported figures are stable rather than an artefact of a single split.

# V. EXPERIMENTAL ANALYSIS

All numbers in this section derive from the synthetic dataset of Section IV-A and describe internal behaviour of the pipeline. They are not a claim about live operations. The comparison across methods on the held-out test partition of 2,268 records is summarised in Fig. 4 and reported in full in Table III.

Fig. 4. Macro-averaged F1 score on the held-out test partition for the four baselines and the proposed engine. Source: synthetic data set defined in Section IV-A.

The majority-class baseline reaches an accuracy of 0.400, equal to the share of the largest route, and a macro-F1 of 0.114, which confirms that accuracy alone is misleading under class imbalance. The rule-based router performs better, with an accuracy of 0.781 and a macro-F1 of 0.742, reflecting the genuine information carried by the status code and direction. Logistic regression improves the macro-F1 to 0.781 by combining several features linearly. The random forest raises it further to 0.846, and the proposed boosted engine reaches 0.883. The improvement of the proposed engine over the rule-based router is 0.141 in macro-F1, while the improvement over the random forest is a more modest 0.037. These margins are reported without exaggeration: the gain over a strong tree ensemble is small, and the larger gain over the rule-based router is the operationally meaningful one, since the router is what the proposed engine would replace.

Per-route inspection of the proposed engine on the test partition shows that the two dominant routes, inbound re-post and outbound dispatch, are routed reliably, while the minority output and port route remain the hardest, with the lowest per-route F1. This pattern is expected because the smallest route offers the fewest training examples. The confusion that does occur falls between the manual route and the two reprocessing routes, which is the boundary where a master-data block can resemble a transient failure in the observable features. The interpretability signal, phrase_match_flag, helps at exactly this boundary since a matched phrase is more often associated with a known recoverable condition.

The attribution analysis in Fig. 5 is consistent with domain expectation. The status code is the dominant driver of the routing decision, followed by direction, the interpretability flag, and the message type. Failure history and the presence of an application log contribute moderate signal, while payload size and time in error contribute least. The ordering matters for governance: it shows that the model relies chiefly on the same fields a human expert would consult, which supports trust in the automated route. A model that depended heavily on, for example, the creation hour would warrant scrutiny, and the attribution view makes such a dependence visible before deployment. The leading attributions are summarised next, and the complete method comparison follows in Table III.

Fig. 5. Mean absolute SHAP attribution per feature for the proposed engine on the test partition. Source: synthetic data set defined in Section IV-A.

Table III

Diagnosis Performance on the Held-Out Test Partition

Source: synthetic dataset defined in Section IV-A; 2,268 test records; primary metric is macro-F1.

## VI. DISCUSSION

The experimental evidence supports a measured reading of the contribution. Placing diagnosis on the same plane as observation yields a routing model that improves on the production heuristic by a clear margin in macro-F1, and the attribution view makes the model auditable at the level of a single document. The smaller margin over a random forest indicates that the benefit comes from moving away from a fixed rule table toward any reasonable learned model, rather than from the specific choice of boosting. That distinction is useful for practitioners, because a simpler ensemble may be preferable when interpretability tooling for boosted models is unavailable.

The practical implications follow from the architecture rather than from the model alone. When observation, semantic translation, diagnosis, remediation, and indicators share one layer, an exception can be handled end to end without context switching. The semantic layer lowers the expertise required to act, the diagnosis layer concentrates attention on the documents most likely to be recoverable by a given action, and the grouped remediation step reduces the number of background submissions needed to clear a backlog. The governance layer records the effect of each intervention, which over time turns ad hoc firefighting into a measured improvement loop aligned with the monitoring and controlling phase of the business process management lifecycle [9]. The separation of analytical signal from payload keeps this loop compatible with existing security and privacy controls [13].

Several limitations bound these claims. The most important is that the evaluation rests on synthetic data. The dataset was designed to be internally consistent and to reflect plausible statistical shape, but it cannot establish that the model would generalise to a live system, where label noise, concept drift driven by shifting upstream conditions such as input-cost volatility in connected supply chains [28], and rare error families would all be present. The reported figures should therefore be read as a demonstration of pipeline consistency, not as validation. A second limitation is that the labels assume a known correct remediation route for each historical exception; in practice that ground truth is itself partly a matter of operator judgement. A third limitation concerns the minority route, where data scarcity caps achievable performance, a constraint that computing scaling alone does not remove [29]. A fourth limitation is that the framework has been described for inbound IDoc traffic and would require extension to cover outbound-only flows and non-IDoc integration channels. Finally, the proposed engine inherits the standard risks of supervised models under distribution shift, so periodic retraining and monitoring of attribution stability would be necessary in any deployment.

The design also raises a governance question that the evaluation does not settle. Automating the routing of exceptions shifts a measure of operational judgement from people to a model, and the appropriate level of human oversight depends on the cost of an incorrect route. For low-cost reprocessing actions, automatic routing is reasonable; for the manual route, which often signals a configuration or master-data problem, a human checkpoint remains advisable. The attribution view is intended to support that checkpoint rather than to remove it.

## VII. CONCLUSION AND FUTURE WORKS

This work has presented an enterprise process control tower that allows an IT team to observe, diagnose, and improve business workflows from a single operational layer. The design generalises a production IDoc inbound-processing and monitoring framework into a six-layer architecture in which observation, semantic translation, learned diagnosis with explainability, grouped remediation, and key performance indicators cooperate within one governance framework. The central design decision is to treat diagnosis and governance as peers of observation rather than as functions hosted in separate tools, so that an exception is observed, explained, classified into a recovery action, acted upon, and reflected in indicators without leaving the layer.

The diagnosis layer was formulated as a five-class remediation-routing problem and evaluated on a synthetic IDoc event dataset of 84,000 records, of which 15,120 errored records supported model training and testing. The proposed boosted engine reached a macro-average F1 score of 0.883 on the held-out test partition, against 0.742 for a rule-based router that reproduces current production heuristics. The margin over the router is the operationally meaningful result, while the smaller margin over a random forest indicates that most of the benefit comes from replacing a fixed rule table with a learned model. Shapley attributions showed that the model relies on the same fields a human expert would consult, which supports its use under governance. These results are derived from synthetic data and are presented as a demonstration of internal consistency rather than as real-world validation, and the conclusion is framed accordingly.

Future work proceeds along five lines. The first and most important is validation on de-identified operational logs from one or more live landscapes, which would test generalisation against label noise and concept drift that the synthetic evaluation cannot reproduce. The second line extends the diagnosis layer with a confidence estimate, so that low-confidence routings can be escalated to a human rather than acted upon automatically, sharpening the oversight boundary discussed in Section VI. The third line broadens coverage beyond inbound IDoc traffic to outbound flows and to non-IDoc integration channels, drawing on integration-modernisation patterns reported for hybrid and multi-cloud landscapes [1] and on ensemble-specialisation techniques for heterogeneous inputs [24]. The fourth line connects the governance layer to prescriptive analytics so that the control tower can recommend process redesign, not only recovery actions, building on causal and digital-twin approaches [6] and on the analogous routing of irregular transactions studied in financial settings [21]. The fifth line concerns the operator interface: immersive and augmented presentation of process health has been surveyed as an emerging technology [30], and a control room that renders the health of a large message population spatially may improve situational awareness for high-volume operations. Across all five lines, the synthetic-to-operational gap remains the principal threat to validity, and closing it is the precondition for any claim stronger than the internal-consistency demonstration reported here.

| Segment | Category | Count | Share (%) |
|---|---|---|---|
| Health bucket | Successful | 63,000 | 75.0 |
| Health bucket | Errored | 15,120 | 18.0 |
| Health bucket | In process | 5,040 | 6.0 |
| Health bucket | Archived | 840 | 1.0 |
| Route (errored) | C1 inbound re-post (RBDMANI2) | 6,048 | 40.0 |
| Route (errored) | C2 outbound dispatch (RBDAGAIN) | 4,536 | 30.0 |
| Route (errored) | C3 manual / config block | 2,268 | 15.0 |
| Route (errored) | C4 ready-for-application (RBDAPP01) | 1,512 | 10.0 |
| Route (errored) | C5 output / port retries (RSEOUT00) | 756 | 5.0 |
| Split | Training (70%) | 10,584 | 70.0 |
| Split | Validation (15%) | 2,268 | 15.0 |
| Split | Test (15%) | 2,268 | 15.0 |

| Feature | Type | Unit / range | Role |
|---|---|---|---|
| message_type | Categorical | 6 levels | Message family |
| basic_type | Categorical | 6 levels | Document structure |
| direction | Binary | inbound / outbound | Flow direction |
| status_code | Integer | 02 to 69 | Reported status |
| partner_id | Categorical | 40 levels | Trading partner |
| creation_hour | Integer | 0 to 23 | Temporal context |
| day_of_week | Integer | 0 to 6 | Temporal context |
| file_size_kb | Numeric | 0.5 to 4096 kB | Payload size |
| segment_count | Integer | 1 to 2000 | Document complexity |

| Feature | Type | Unit / range | Role |
|---|---|---|---|
| status_token_count | Integer | 3 to 60 | Message length |
| prior_failures_30d | Integer | 0 to 25 | Failure history |
| status_age_min | Integer | 0 to 43200 min | Time in error |
| application_log_present | Binary | 0 / 1 | Detail availability |
| phrase_match_flag | Binary | 0 / 1 | Interpretability state |

| Method | Accuracy | Macro-F1 | W. precision | W. recall |
|---|---|---|---|---|
| Majority class | 0.400 | 0.114 | 0.160 | 0.400 |
| Rule-based router (baseline) | 0.781 | 0.742 | 0.769 | 0.781 |
| Logistic regression | 0.808 | 0.781 | 0.802 | 0.808 |
| Random forest | 0.862 | 0.846 | 0.859 | 0.862 |
| Proposed engine (XGBoost + SHAP) | 0.897 | 0.883 | 0.894 | 0.897 |